\documentclass[conference]{IEEEtran}

\pdfoutput=1

\usepackage{cite}

\usepackage{amssymb}
\usepackage{amsmath}
\usepackage{amsfonts}

\usepackage{float}
\usepackage{newfloat}
\usepackage{adjustbox}
\usepackage{cite}
\usepackage{subfig}

\newcommand{\mb}{\mathbf}
\newcommand{\mc}{\mathcal}

\newcommand{\tb}{\textbf}
\newcommand{\tbu}[1]{\textbf{\underline{#1}}}

\newcommand{\ndn}[1]{nodeN$_{#1}$}

\DeclareGraphicsExtensions{.eps,.jpg,.gif,.png,.pdf}
\graphicspath{{figs/}}

\ifCLASSINFOpdf
\else
\fi

\usepackage{url}

\usepackage{makeidx}  
\usepackage{enumitem}

\DeclareFloatingEnvironment[name=Protocol,placement=htb,within=none]{protocol}
\newenvironment{tprotocol}
{   \setlength\intextsep{\baselineskip}
    \setlength\floatsep{\baselineskip}
        \setlength\textfloatsep{\baselineskip}

    \begin{protocol}
    \begin{adjustbox}{minipage=.97\linewidth,fbox,center}
}{\end{adjustbox}\end{protocol}}

\newenvironment{tenum}{
\begin{enumerate}[align=left, leftmargin=.15cm]

}{\end{enumerate}}

\begin{document}

\title{Probing Attacks on Physical Layer Key Agreement for Automotive Controller Area Networks}

\author{\IEEEauthorblockN{Shalabh Jain$^1$, 
Qian Wang$^2$, 
Md Tanvir Arafin$^2$ and 
Jorge Guajardo$^1$}
\IEEEauthorblockA{$^1$ Robert Bosch LLC, Research and Technology Center, Pittsburgh, PA 15222, USA \\
Email: \{shalabh.jain, jorge.guajardomerchan\}@us.bosch.com}
\IEEEauthorblockA{$^2$ Department of Electrical Engineering, University of Maryland, College Park, MD 20742, USA \\
Email: \{qwang126, marafin\}@umd.edu}}

\maketitle

\footnotetext[2]{Work while the authors were at Bosch Research and Technology Center, Pittsburgh, PA}

\begin{abstract}
Efficient key management for automotive networks (CAN) is a critical element, governing the adoption of security in the next generation of vehicles. A recent promising approach for dynamic key agreement between groups of nodes, \emph{Plug-and-Secure} for CAN, has been demonstrated to be \emph{information  theoretically} secure based on the physical properties of the CAN bus. In this paper, we illustrate side-channel attacks, leading to nearly-complete leakage of the secret key bits, by an adversary that is capable of probing the CAN bus. We identify the fundamental characteristics that lead to such attacks and propose techniques to minimize the information leakage at the hardware, controller and system levels.
\end{abstract}

\IEEEpeerreviewmaketitle

\section{Introduction}%
\label{sec:introduction}%
Over the past few years, an increase in the number of connectivity interfaces on a traditional automobile has drawn the attention of the security community. Several high profile attacks have been demonstrated on the modern car by academic and industrial researchers, e.g. \cite {hoppe_2008_autothreats,kohno_2011_car_attack_surface,kohno_2010_carsec,%
miller_2015_jeep,trappe_2010_tpms_security}.
These attacks are primarily facilitated by the lack of security (authentication, encryption) in the existing architecture of the controller area network (CAN). Concurrently, several techniques have been proposed for integrating security into the current architecture, \cite{jorge_2012_auto_sec,ingrid_2011_canauth,andre_2014_automotive,intel_2015_autosec}.

Similar to traditional secure systems, several of these techniques rely on secure provisioning of symmetric keys within the nodes on the CAN bus. However, secure and robust provisioning, maintenance and update of cryptographic keys within the automotive supply chain can be incur significant overhead, and may require changes to the automotive manufacturing and servicing facilities. Several commercial systems to enable such a process have been proposed, e.g. in \cite{etas_2016_kms}.

Dynamic generation and distribution of keys in a secure manner can provide an alternative (or reduce the functional requirements) of traditional provisioning systems. One promising approach, \emph{Plug-and-Secure} for CAN (PnS-CAN), has been recently proposed towards this goal, in \cite{andreas_2015_pns_basic_conf, jain_jorge_2016_ches}. A key advantage of the \emph{Plug-and-Secure} scheme is the utilization of inherent physical layer properties of the CAN bus to provide security guarantees for key agreement between groups of nodes.

Cryptographic systems that are provably secure in the computational model have often been compromised by exploiting characteristics of their physical implementation. Physical characteristics such as timing differences, power leakage, and other features or radiation can provide a covert communication medium, \emph{side-channels}, leaking system information to an adversary. Several attacks have been demonstrated on traditional systems using side channels, e.g. in \cite{kocher_96_timining_rsa, kocher_99_dpa, bernstein_2005_cache_sidec, jean_pierre_2006_bp_sidec}. Comprehensive analysis of a system to identify and exploit such side-channels can be difficult. It has been observed from traditional systems that prevention of such attacks can add significant overhead to system design and negatively impact performance.

In this paper, we demonstrate that other properties of physical layer the CAN bus can be utilized to violate the security of the PnS-CAN system by an adversary capable of probing the CAN bus. Here we discuss several \emph{voltage}, \emph{timing} and \emph{transient characteristics} based side-channels that can be used to attack the system and partially extract the secret keys. We then propose countermeasures that can be implemented at different levels of the system, namely low level hardware changes, controller modifications and system level changes. Our methods align well with existing system configuration and result in little overhead.

\subsection{Our Contributions}
We investigate the side-channels for the two-party PnS-CAN protocol proposed in \cite{andreas_2015_pns_basic_conf}. Our contributions are as follows,
\begin{itemize}
\item We identify characteristics of the CAN bus that can be utilized to extract the secret key during execution of the PnS-CAN protocol. We outline a general methodology to map existing CAN identification techniques to attack our scheme. We further demonstrate the attack of the system due to one of the non-trivial feature, i.e. timing characteristics.
\item We propose countermeasures that can be included in the transceiver hardware or the CAN controller to minimize such leakage.
\item For key agreement between groups using the PnS-CAN protocol, we propose system level changes that can be used to further reduce the side-channel leakage.
\end{itemize}

\subsection{Related Work}
\noindent \textbf{Side-channels:} Since the initial results of Kocher \cite{kocher_96_timining_rsa}, a lineage of work has been created to utilize the state based differences in a physically observable phenomenon, such as timing, power, etc, to estimate the secret state. Several attacks have been demonstrated on traditional systems using timing characteristics \cite{kocher_96_timining_rsa, brumley_2003_remote_timing}, power characteristics \cite{kocher_99_dpa}, and a variety of other side-channels \cite{bernstein_2005_cache_sidec, jean_pierre_2006_bp_sidec}. Our results extend this line of work and utilize the differences in the physical characteristics between various participating nodes to extract the secret key. To the best of our knowledge, there is no work in literature to identify low-level attacks on the PnS-CAN scheme.

A critical difference of our work from existing side-channel literature is that traditionally, side-channel attacks extract a pre-existing secret key based on usage of the key. In our work however, we extract the secret key during its \emph{derivation} phase, i.e. prior to storage. Thus, existing methods can be applied in conjunction with our work to extract any secret bits that remain.
\\

\noindent \textbf{Intrusion detection systems (IDS) based on CAN signal characteristics:} The lack of node-identity and authentication in the traditional CAN protocol has led several researchers to utilize physical signal (or device) characteristics as a fingerprint to identify the transmitting nodes. These results, e.g. in \cite{shin_2016_ecu_fingerprint_timing, groza_2014_ecu_id, btp_2016_ml_ecu}, highlight the existence of subtle, yet identifiable differences between the transmissions from different nodes on the CAN bus. Though such features have been explored and utilized to obtain very robust identification features in traditional network domains, i.e. wireless networks, Ethernet based networks or aquatic networks, they are in very early stages of investigation for CAN.

Since the PnS-CAN system relies on the inability of an adversary to identify the transmitter, any fingerprint associated with the transmitter of a node can be used to attack PnS-CAN. It should be noted that these results were investigated in context of increasing system security by designing an IDS based on the identified features. However, in our case, these features are utilized to attack the system.

The authors in \cite{groza_2014_ecu_id} use simple features such as mean-square error between bit samples, and convolution amplitude to fingerprint and identify the ECUs. However, the performance of these mechanisms is highly dependent on the message value, which can vary significantly for our schemes. This was improved in \cite{shin_2016_ecu_fingerprint_timing} by using several time domain and frequency domain features of the transmitted signals. The authors in \cite{btp_2016_ml_ecu} further improve the results and utilize a variety of classifiers, ranging from simple binary support vector machines (SVM) to multi-layered convolutional neural network (CNN) to identify the ECUs. They demonstrate that despite using low precision hardware for sensing, they can achieve improvement over traditional classifiers by utilizing powerful methods such as CNN. Intuitively, these identification mechanisms can be utilized to break the PnS system. There are however several challenges and simplifications that additionally apply to the PnS system. For example, these systems rely on classification of a message frame consisting of a single transmitter and synchronization source. However, for the PnS-CAN system, this assumption does not apply directly. On the other hand, simpler features may be used due to a slightly different structure of the problem.

\subsection{Organization}
The remainder of the paper is organized as follows. In Section \ref{sec:prelim}, we describe the basic PnS-CAN system and the adversarial model. We provide an overview of the useful physical characteristics in Section \ref{sec:attack_overview}, and demonstrate an attack on an unprotected system. We propose several countermeasures to prevent these attacks in Section \ref{sec:countermeasure}. In Section \ref{sec:discussion}, we discuss the practical applicability of our countermeasures and future directions.

\section{Preliminaries}
\label{sec:prelim}

\subsection{Notation}
\label{sec:notation}
We adhere to the following notation for the paper. For two nodes \ndn 1, \ndn 2 executing the protocol, we denote by $x-y$, the simultaneous transmission of $x$ by the primary node, i.e. \ndn 1 and $y$ by \ndn 2. Here, $a,b$ are logical bits, i.e. $x, y \in \{0,1\}$. We denote by $GW$, the central entity that controls (or initiates) the execution of the key agreement protocol. Such a functionality is typically performed by the central gateway in the network.

\subsection{CAN Bus Physical Layer}
\label{sec:can_bus_prelim}
CAN bus, the primary communication network for most modern day cars, is a broadcast medium consisting of a series of nodes connected via a twisted-pair cable with termination impedance at either end. It has two logical states, the dominant `0' state, where the bus is driven by a voltage, and the recessive `1' state, where the bus is grounded. If two nodes transmit a bit simultaneously, the effective state of the bus is dominant `0' if any of the nodes transmits the dominant bit. Thus, the bus acts as a logical AND gate between inputs from the nodes.

The CAN bus utilizes differential signaling to transmit the data. In the CAN standard, when transmitting the dominant bit $0$ on the bus, the output pins of the nodes, CANH and CANL, are driven to different voltage levels, and the difference from CANH to CANL is the output of the CAN bus. Similarly, transmission of a recessive bit $1$ occurs when CANH and CANL are not driven, and will have similar voltage levels.

\subsection{System Model}
\label{sec:sys_model}
As this work focuses on attacking the PnS-CAN protocol, it inherits the system requirements for successfully building that system, as enumerated in \cite{jain_jorge_2016_ches}. We expect the typical automotive network (CAN) to be comprised of heterogeneous nodes, i.e. nodes from different manufacturers or families. Since the PnS-CAN protocols are based on simultaneous transmission by two nodes, all write operations on the bus during the key agreement phase use ECU pairs. Thus, the ECUs require modified CAN controllers that allow simultaneous transmission of data during this phase. However, during regular operation, the network operates in a CAN-compliant manner, and hence has only single node transmitting during a frame.

\subsection{Adversarial Model}
\label{sec:adv_model}
\begin{figure}[!tb]
\centering
\includegraphics[width=3.5in]{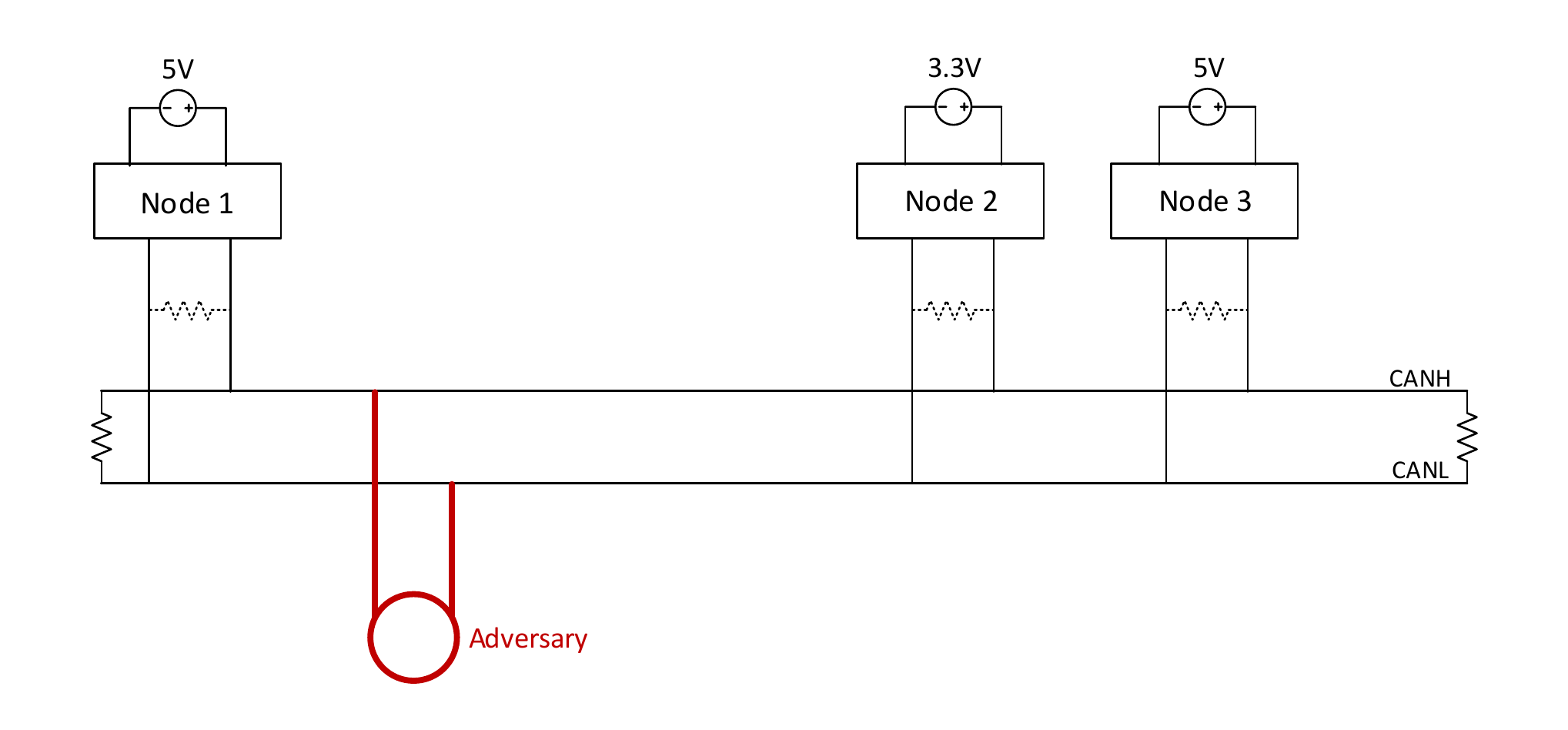}
\caption{Representation of a typical CAN bus with $3$ nodes and an eavesdropping adversary}
\label{fig:basic_can_bus}
\end{figure}

We consider an arbitrarily powerful adversary that is capable of observing the variation of CAN bus signals with high voltage precision and timing resolution. Such an adversary can be simply realized by an eavesdropper who accesses the wires directly using a high precision oscilloscope. An alternate means could be through a regular ECU connected to the CAN bus with a high precision analog-to-digital (A/D) converter at the input and a modified CAN controller capable of sampling the bus at a high frequency. In a car, such nodes can be connected to the OBD-II diagnostics port. A representation of the CAN bus with adversarial presence is illustrated in Fig. \ref{fig:basic_can_bus}.

The goal of this work is to demonstrate the existence of side-channels. Even though such powerful adversaries are capable of actively injecting or modifying signals, we consider the adversarial behavior to be restricted to \emph{passive} observations. However, we assume that the adversary is capable of observing the system for an arbitrarily long time. During regular system operation, we assume that the adversary is capable triggering the ECU of its choice to transmit. However, we assume that such selective modifications cannot be made during execution of the PnS-CAN protocol. Further, we assume that the adversary cannot control the content of the transmissions.

For this paper, we limit the adversarial access to a single point on the network. It should be noted that multiple points of observation can undoubtedly increase the leakage. However, since in an unmodified system, a single adversary is sufficient for complete compromise, we leave the consideration of multiple probes to future analysis. For our attacks, we assume the location of the adversary to be carefully selected and fixed. An adversary can theoretically improve its observations by adaptively changing its observation point.

\subsection{Experimental Setup}
\label{sec:exp_setup}
Our experimental setup utilizes an FPGA (Altera Cyclone-V) based implementation of PnS-TwoParty scheme described in Protocol \ref{prot:two_party_pns}, implemented into a modified CAN controller. The modified controller generates fully compliant CAN2.0 frames that can be accepted by any traditional CAN controller. The test network consists of 16 nodes, using custom designed transceiver boards that utilize the FPGA digital outputs, connected via the standard CAN twisted pair cable.

The typical CAN architecture, consists of several subnetworks of ECUs connected by one or more powerful nodes that act as gateway nodes ($GW$). However, for our system, the nodes are connected in a single chain with the $GW$ at one end. This setup closely emulates one subnet of the network found in most modern cars. We use a commercial USB-CAN module to monitor the bus traffic. We use an Agilent 6012A oscilloscope to probe the bus and record the samples for offline processing.

\subsection{PnS-CAN Scheme}
\label{sec:pns_scheme}
The PnS-CAN scheme described in \cite{andreas_2015_pns_basic_conf, jain_jorge_2016_ches},  enables key agreement between multiple nodes connected to the CAN bus. For completeness, here we present a brief overview of the fundamental operations of the two-party and linear group-key schemes from \cite{jain_jorge_2016_ches}. We include specific implementation aspects, that aid our attacks, in the protocol definition.  For detailed discussion about the security aspects of the original scheme, the reader is referred to \cite{jain_jorge_2016_ches}.

The PnS-CAN scheme between two nodes utilizes the wired AND property of the CAN bus to mask the bits simultaneously transmitted by the nodes. The security of this scheme is based on the inability of an eavesdropper to differentiate between transmissions that result in the same logical output on the bus, i.e. combinations that result in the dominant ($0$) output.

\begin{tprotocol}

Protocol 1: \tb{PnS-TwoParty(\ndn 1, \ndn 2, $r_1$, $r_2$)}
\begin{tenum}
\item \ndn 1 generates a sequence of psueodrandom bits, $s_1$ using seed $r_1$. The sequence is split into chunks ($c_1(i)$) of half the size of the maximum CAN 2.0 payload. i.e. $s_1 = \{c_1(1) || c_1(2) || \ldots\}$. Similarly, \ndn 2 generates a sequence of psueodrandom bits, $s_2 = \{c_2(1) || c_2(2) || \ldots\}$, using seed $r_2$.
\item The bits in each chunk are interleaved with their complements to produce packets of size equal to the maximum payload, namely $\{p_1(1), p_1(2), \ldots\}$ and $\{p_2(1), p_2(2), \ldots\}$. Note for interleaving, each $0$ bit is replaced by $01$ and $1$ is replaced by $10$.
\item \label{item:pns2:tx} \ndn 1 initiates the transmission by using a dedicated PnS-CAN header and $p_1(1)$ as payload. \ndn 2 synchronizes to the header and simultaneously transmits packet $p_2(1)$ during the payload phase. For compliance with regular CAN frames, each node dynamically performs bit stuffing and CRC computation based on the bus outputs.
\item \ndn 1 and \ndn 2 sample the first logical bus output sequence as b(1). The bits are de-interleaved to identify the the result of the original and complement transmission. Location within the transmitted chunk, where both the non-inverted and inverted bits produced a $0$ in the bus output, are secret from an ideal eavesdropper. These are considered as secret bits and the remaining bits are discarded.
\item The resulting sequence at \ndn 1(remaining bits in the chunk) are always inverse of the sequence at \ndn 1. Thus \ndn 2 inverts its sequence to obtain a sequence of secret bits identical to \ndn 1. The process is repeated from \ref{item:pns2:tx} until the desired number of secret bits are generated.
\end{tenum}
\caption{Practical two-party PnS protocol}
\label{prot:two_party_pns}
\end{tprotocol}

The operation of the basic two party protocol between \ndn 1 and \ndn 2, using random seeds $r_1, r_2$ is illustrated as Protocol \ref{prot:two_party_pns}. In this, a secret bit is generated between \ndn 1 and \ndn 2  when one of the nodes transmits a logical $0$ (dominant bit) while the other transmits a logical $1$ (recessive bit). Note that an ideal adversary is unable to identify which of the two nodes, \ndn 1 or \ndn 2 transmitted the dominant bit.
The group key scheme described in \cite{jain_jorge_2016_ches} utilizes the PnS-TwoParty protocol between successive nodes to form a PnS chain. It was demonstrated that pairwise interactions between consecutive ECUs are sufficient to share the key with the group.
Consider a group $G_K = \{N_1, N_2, \ldots, N_K\}$ of $K$ nodes. The group key scheme can be described as
\\

\noindent \underline{Protocol 2: \tb{GroupKey-Linear:}}
\begin{enumerate}
\item The group communication order $G_K = \{N_1, N_2, \ldots, N_K\}$  is broadcast to all participants by the central gateway ($GW$).
\item First, \ndn 1 performs PnS-TwoParty with \ndn 2 to establish a key $k_{12}$.
\item Then \ndn 2 utilizes the result $k_{12}$ as the seed to generate its random sequences and performs PnS-TwoParty with \ndn 3.
\item The process continues till \ndn K is reached. The final key is the group key shared between all nodes.
\end{enumerate}

It can be observed, that the security of the group key protocol depends on the security of the pairwise PnS-TwoParty scheme. We note that in systems, where the adversaries only have high-level (software) access to the nodes, they can only observe the logical output of the bus as determined by a single (or three) transceiver samples. Thus, though perfect obfuscation of the bit values in PnS-TwoParty is theoretically possible, it cannot be guaranteed for most practical systems where the adversaries may observe multiple high-resolution bus samples.

\section{Attacks on PnS-CAN}
\label{sec:attack_overview}
To illustrate the properties that can be exploited to mount a successful attack, consider the PnS-TwoParty system. As described in Section \ref{sec:pns_scheme}, a secret bit results when one of the nodes transmits the dominant bit $0$, i.e. drives the bus, and the other node transmits the recessive bit $1$, i.e. performs no action. Even though, in the PnS-CAN system, nodes transmit messages as full frames, for each bit of significance (secret bit), only a \emph{single} node is driving the bus. Thus identification of the transmitting node effectively leaks the bit, as the bit is 0 if the driving node is primary participant and 1 otherwise.

For example, consider a PnS interaction between \ndn 1 and \ndn 2 using the \emph{random} sequences $\{0, 1, 1, 0\}$ and $\{1, 0, 0, 1\}$ respectively. This results in $4$ shared secret bits, i.e. key = $0110$. The bit observations on the bus corresponding to these would comprise of 8 bits (random bits interleaved with the complement). $\{ (b_1, b_2), \ldots, (b_7, b_8),~b_i = 0$. Here, $b_1$ results from \ndn 1 transmitting a dominant bit and \ndn 2 a recessive bit. Thus if the adversary can identify \ndn 1 as the active node during $b_1$, it can learn that the first secret bit is $0$. Next, we describe the phenomenon that can be used to differentiate between various transmitters based on physical properties of the CAN bus.

\subsection{Physical Characteristics}
\label{sec:phy_char}
Similar to other electrical systems, in an automotive network will have differences in characteristics of the driver and network between an observer and the transmitters. We illustrate 3 phenomena that can be utilized to differentiate between bits transmitted by different nodes.
\\

\noindent \tb{Steady state characteristics} - One of the advantages of using differential signaling for CAN transmissions, is that it enables devices with varying electrical characteristics to be utilized on the same bus without any additional compensation circuitry. While this improves the design and robustness of the bus, the different characteristics also enables= identification of the transmitter, leading to leakage of the bits. For a transmitted bit, differences in steady state characteristics can be attributed to several factors,

\begin{figure*}[!tb]
\centering
     \subfloat[Observations for scenario of 2 nodes, A:MCP2551, B:TJA1040\label{fig:s_vth_d_manu}]{%
       \includegraphics[width=3in]{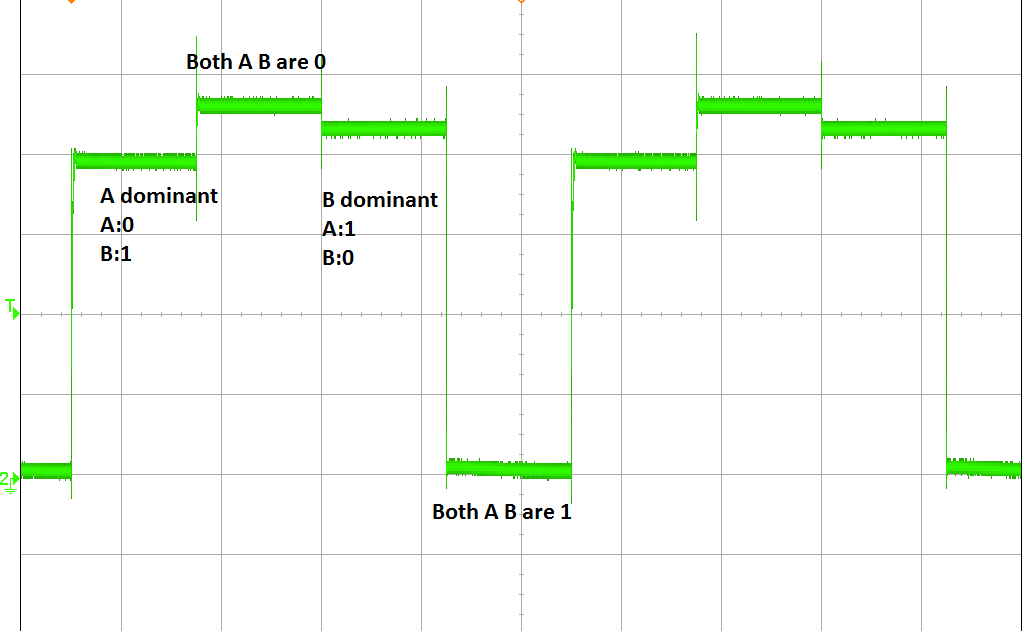}
     }\hfill%
     \subfloat[Observations of two identical transmitters located at different distances from the adversary\label{fig:s_vth_s_manu_imp}]{%
       \includegraphics[width=3in]{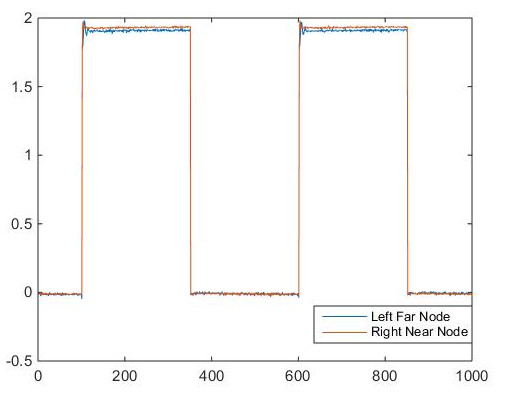}
     }
     \caption{Steady state characteristics}
     \label{fig:ss_char}
\end{figure*}

\begin{enumerate}[align=left, leftmargin=.15cm]
\item Difference in driver circuits - Transceivers from different manufacturers (or even different models of the same manufacturer) can have different drive characteristics and output voltage range of the CANH and CANL pins. This can be due to different circuits, components or load impedance. Thus an adversary measuring the absolute voltage on CANH and CANL lines with respect to a common ground reference can distinguish between the dominant transmissions from different nodes.

For example, consider a realization of the network in Fig. \ref{fig:basic_can_bus} using Microchip (MCP2551) for \ndn 1 and NXP (TJA1040) for \ndn 2. The ideal specified range for the CANH pin for MCP2551 is between $2.75V$ and $4.5V$, while the same range for TJA1040 is between $3V$ and $4.25V$. In Fig. \ref{fig:s_vth_d_manu}, we illustrate the voltage observations of the adversary for a sequence of bits corresponding to the generation of a secret bit in the PnS-CAN protocol, i.e. transition from a $0-1$ scenario to a $1-0$ scenario. The adversary can clearly distinguish between the dominant transmission by \ndn 1 and \ndn 2.
\\
\item Different operating voltages - The differential signaling of the CAN bus allows interoperability of nodes with different supply voltage, without scaling circuits. For example, typical automotive networks contain ECUs with both 5V and 3.3V operating voltage. Though these nodes have similar differential voltage between the CANH and CANL lines, the absolute voltage level on each line, during bit transmission, is different. An adversary can utilize this identify the node transmitting the dominant bit during PnS-CAN.
\\
\item Different physical locations - Even nodes with identical drivers and operating voltages can seem different from the view of an intermediate observer in the network. This is due to the differences in the effective impedance of the network segment between the two transceivers and the observer point. For a typical CAN bus scenario, several factors can contribute to these differences, e.g. different length of wires between the nodes and the observer, or different number of intermediate nodes. In Fig. \ref{fig:s_vth_s_manu_imp}, we illustrate the differential voltage of two nodes at varying distances (over $1 m$ difference) from the observer. In a typical CAN network, the difference in distance can be over $30 m$Though this difference appears small in comparison to other phenomenon, it can be useful in many scenarios

\end{enumerate}

\noindent \tb{Transient characteristics} - The CAN physical medium has a non-negligible capacitance and inductance that influences the signal as it propagates. This influence, for signals transmitted by different nodes, is in general non-uniform. Coupled with inductive-loading of different intermediate transceivers between an observer and the nodes, these influences can have observable impact on the signals. Thus, as the state of a node transitions from recessive to dominant (or vice versa), it exhibits different transient characteristics. The sample point of a typical bit is sufficiently delayed to ensure that the CAN bus is robust from such transient fluctuations. However, an adversary observing with a higher resolution can use these differences to identify the transmitter.

\begin{figure*}[!htbp]
     \subfloat[Observations of transients by an adversary, centrally located between two identical transmitters\label{fig:transient_2node}]{%
       \includegraphics[width=3in]{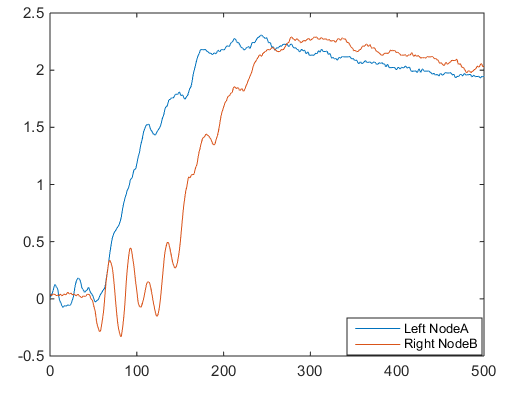}
     }\hfill%
     \subfloat[Timing observations of an adversary centrally located between two transmitters\label{fig:adv_002pc}]{%
       \includegraphics[width=3in]{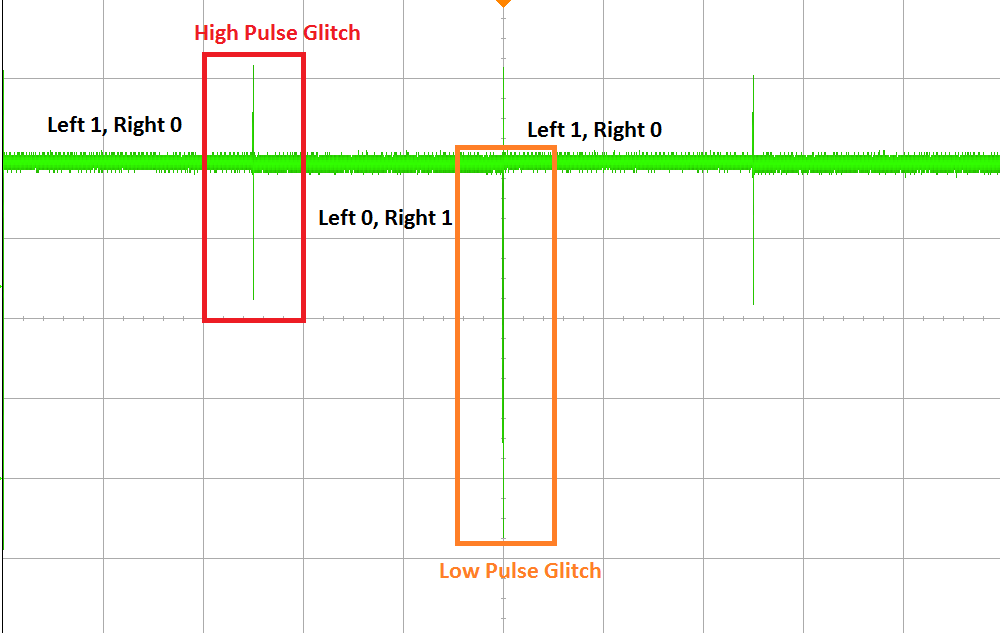}
     }
     \caption{Transient and timing characteristics}
     \label{fig:trans_timing_char}
\end{figure*}

One of the benefits of transient characteristics is that they can vary widely even for similar nodes placed symmetrically with respect to an observer. As illustrated in Fig. \ref{fig:transient_2node}, observations by an adversary that is symmetrically placed in Fig. \ref{fig:basic_can_bus} relative to two identical transmitters can be clearly distinguished, based on time domain characteristics. However, a disadvantage of the transient characteristics is that they cannot be compared directly, and the performance is highly dependent on the features that are derived for identification.  Several such features have been enumerated in \cite{shin_2016_ecu_fingerprint_timing}. \\

\noindent \tb{Timing characteristics} - The typical propagation delay for the twisted pair cable used as the physical medium for the CAN bus is $5ns/m$. Thus for a traditional network of length upto $50m$, the difference in the time the transmitter drives (or releases) the bus and an observer observes a signal transition can be up to $250ns$. Though such delays are accommodated within the CAN bit timing specification for correct sampling of the bit value, they can be exploited by an adversary to identify the transmitter of a bit.

The relative bit timing observed by an adversary for two transmitters can be influenced by two factors, namely the difference in propagation delay between the observer and the transmitters and the synchronization offset between the transmitters. In Fig. \ref{fig:adv_002pc}, we illustrate the observations of an adversary that is symmetrically placed relative to two identical transmitters. However, as it can be seen, even a propagation difference of $0.02\%$ of the bit timing is sufficient to distinguish between the dominant transmission from the left node and the right node, i.e. the scenario of secret bit generation, $1-0$ followed by $0-1$ transmission in PnS-CAN.

\subsection{Generalized Attack Model}
\label{sec:gen_model}
Each PnS interaction occurs between two nodes in the network. As discussed earlier, the attack on the PnS-CAN system simply reduces to identification of the transmitting node for each individual bit of the frame. Since there are just two nodes in each PnS-CAN round, the adversary simply needs to \emph{distinguish} between the signals transmitted from the two nodes (binary hypothesis testing). Here, we present a general outline of the attack methodology that the an attacker may follow.
\\

\noindent \tbu{(Data Collection)} The adversary can observe the signal transmission on the bus for a long duration prior to the attack. Note that the any CAN-compatible node can synchronize to the transmitter and identify and sample individual bits in a transmitted frame. Consider that the adversary samples each bit $k$ times. We denote these bit observations by $\mc T = \{\mb x_{-\infty}, \ldots, \mb x_0\}$, where $\mb x_i \in \mc X_k = \mathbb{R}^k$ is a vector of samples of the bit from the bus. We assume that the system has $M$ ECUs connected to the bus. Thus the observations $\mb x_i$ correspond to transmissions from any of the $M$ ECUs over a period of time. For traditional CAN networks, such data is not labeled as the packets do not contain identifying information. However, an adversary can generate labels for the data by activating known ECUs and observing their output.
\\

\noindent \tbu{(Classifier Training)} An adversary trains classification functions, $\mc D_{N_i, N_j}: \mc X_k^{n} \rightarrow \{0, \bar{0}\}^n$, for each pair of labels $N_i, N_j$, that estimate the the sequence of transmitters based on the input observations. We note that using $\{0, \bar{0}\}$ represents the process of differentiating between the transmitters, rather than identification. This represents a larger class of classifiers, as any classifier that can correctly assign the labels, can also differentiate between the transmitters. The adversary can train the classification functions directly from $\mc T$, using a variety of supervised or unsupervised techniques, e.g. binary support vector machines (SVM), LSTM,  Convolutional Neural Networks \cite{alex_cnn}.

For scenarios where the node does not have knowledge cannot acquire knowledge of the participating nodes or where the training data is unlabeled, the node may train a generalized decision function $\mc D_{0,\bar{0}}: \mc X_k^n \rightarrow \{0,\bar{0}\}^n$, that uses a  classification function preceded by a selection approach (such as maximum likelihood based estimator) select the best function. Clearly, the performance of a generalized classifier is sub optimal compared to the classifier for a particular pair of nodes. Thus an adversary, without apriori knowledge of the transmitters has a significant disadvantage.
\\

\noindent \tbu{(PnS-CAN compromise)} During PnS-CAN operation, for an adversary to extract the key, it does not require perfect identification of the node identity. Since the PnS system involves just two ECUs at any time, its task is to simply distinguish between the transmissions from the two participating ECUs. Consider the scenario where two nodes $N_i, N_j$ execute the PnS protocol to produce $n$ bits. The adversary observes the $k$ sample values for each of the $n$ bits as $\mb x_i = \{s_1(i), s_2(i), \ldots, s_k(i)\}, 1 \leq i \leq n$ and uses the classifier $\mc D_{N_i, N_j}$ (or general classifier in scenarios where $N_i, N_j$ is unknown) and obtains a sequence of estimates $(\hat{T_1}, \hat{T_2}, \ldots, \hat{T_n}) \in \{0, \bar{0}\}^n$.

For PnS-CAN, classification without correct labeling of the nodes still reveals the complete secret (or its inverse). Thus the disadvantage of the weaker adversary is simply $1$ bit of entropy, which is insignificant and results in violation of system security.
We note that methods from CAN identification literature, e.g. from \cite{kang_2016_ids_dnn_ecu, shin_2016_ecu_fingerprint_timing, groza_2014_ecu_id, btp_2016_ml_ecu}, can be directly mapped to this generalized outline to attack the PnS-CAN scheme. However, the accuracy may be lower since the identification is for a single bit rather than a group of bits (frame).
%

\subsection{Timing Based Attack Evaluation}
\label{sec:timing_attack}
We demonstrate an attack on the PnS-CAN scheme using the timing characteristics described in Section \ref{sec:phy_char}. Our choice of the timing parameter is governed by several factors
\begin{enumerate}
\item Timing represents a phenomenon that is independent of the node characteristics. In fact, it depends only on synchronization and propagation aspects. Thus it can be applied to all networks.
\item Timing attacks demonstrate the vulnerability of the unprotected system against a very simple adversary.
\item The timing component represents a feature that has not been specifically evaluated in \cite{kang_2016_ids_dnn_ecu}. Other features from \cite{kang_2016_ids_dnn_ecu} can be directly applied to attack PnS-CAN via our framework described in Section \ref{sec:gen_model}.
\end{enumerate}

\begin{table*}[htb]\footnotesize
\parbox{.65\linewidth}{
\centering
\begin{tabular}{c|cccc||c|cccc}
\hline
Node & \multicolumn{4}{c}{Delay (ns)} & Node & \multicolumn{4}{c}{Delay (ns)} \\
ID & Min & Max & Mean & Std & ID & Min & Max & Mean & Std \\ \hline
1 & 138 & 166 & 151.8 & 12.6 & 9 & 118 & 154 & 135.2 & 15.7 \\
2 & 140 & 168 & 153.4 & 12.5 & 10 & 118 & 154 & 135.0 & 15.3 \\
3 & 140 & 168 & 153.8 & 12.6 & 11 & 122 & 156 & 139.1 & 14.7 \\
4 & 140 & 172 & 156.2 & 12.9 & 12 & 124 & 158 & 140.9 & 14.6 \\
5 & 130 & 162 & 144.4 & 14.2 & 13 & 116 & 146 & 130.6 & 12.6 \\
6 & 130 & 160 & 144.8 & 14.0 & 14 & 118 & 146 & 131.2 & 12.3 \\
7 & 132 & 164 & 147.1 & 13.7 & 15 & 118 & 152 & 135.4 & 14.8 \\
8 & 136 & 164 & 148.7 & 12.4 & 16 & 122 & 154 & 137.2 & 13.3 \\
\hline
\end{tabular}
\caption{Signal delays between nodes and observer%
\label{tab:prop_delay}}
}\hfill%
\parbox{.30\linewidth}{%
\centering
\begin{tabular}{ccc}
\hline
Interval overlap & $N_1$ & $N_2$ \\ \hline
6.00 & 2 & 13 \\
6.00 & 2 & 14 \\
6.00 & 3 & 13 \\
32.00 & 11 & 12 \\
34.00 & 9 & 15 \\
34.00 & 10 & 15 \\
36.00 & 9 & 10 \\
\hline
\end{tabular}
\caption{Maximum and minimum overlaps%
\label{tab:prop_overlap}}%
}%
\end{table*}

We utilize the oscilloscope to obtain samples of the differential bus voltage at $125 Msamp/s$. The transitions are identified by as points of large change in bus voltage (greater than the CAN trigger) followed by a steady state over atleast half the bit width. We utilize the $50\%$ point of the transition to compute the rise time, fall time and latency.

First, we investigate the separability of the nodes by measuring the propagation delays of the nodes synced with respect to the observation point. In Table \ref{tab:prop_delay}, we enumerate the propagation delays from each node. Intuitively, nodes that have similar propagation delay would be difficult to differentiate (in the perfectly synchronized scenario). Further, in Table \ref{tab:prop_overlap}, we enumerate combinations of nodes have the least (and maximum) overlap. Such nodes pairs correspond to the nodes with the largest (and smallest) adversarial advantage.

\begin{figure*}[!htbp]
\centering
\includegraphics[width=4.5in]{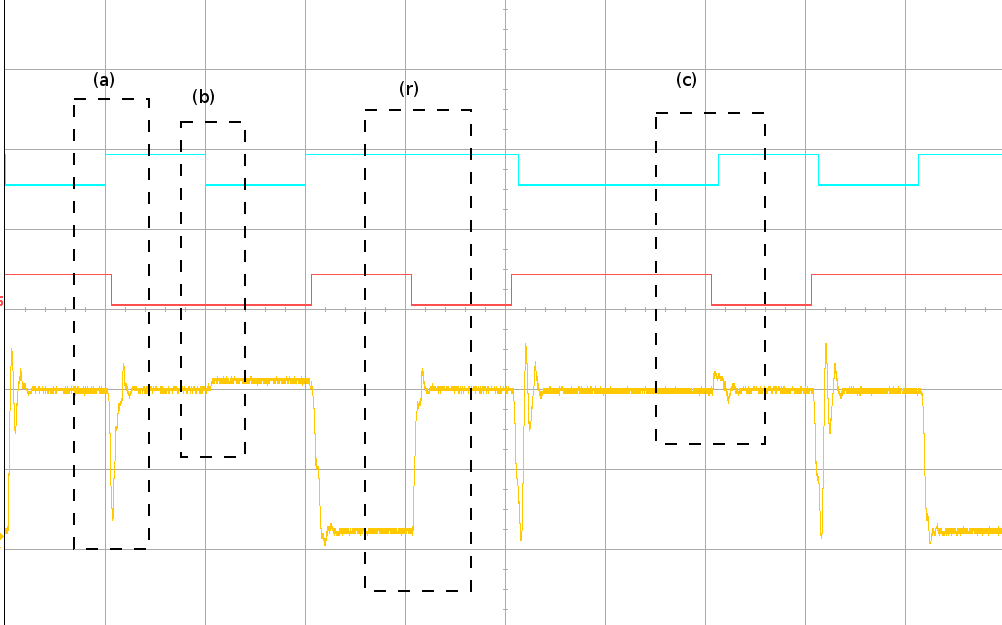}
\caption{Snapshot of bus observations for PnS protocol}
\label{fig:can_timing_attack}
\end{figure*}

In Fig \ref{fig:can_timing_attack}, we illustrate a snapshot of the CAN protocol between two nodes as observed by an adversary. Intuitively, the attack exploits the difference in propagation delays by synchronizing to one of the transceivers and estimating the transmitter based on the rise and fall time offset. There are two key features of the PnS-CAN implementation that  aid our analysis.

\begin{enumerate}
\item The initiating node transmits the PnS header and the secondary node synchronizes to the initiating node. Since only a single node is transmitting during the header phase, it is easy to estimate the timing variation for the synchronous bits.
\item As described in Section \ref{sec:pns_scheme}, the random bits are interleaved with the inverted bits. This introduces dependence in successive transmissions as it enables only certain transitions during the PnS phase. This can be utilized to estimate the bits in some cases.
\end{enumerate}

The detailed attack is be described as \emph{Attack PnS-TwoParty}. We utilized this to identify the secret key for $12$ pairs of nodes in our setup. With minor modification of the threshold parameters between different iterations, we were able to successfully identify all the secret bits exchanged between each pair.
\\

\noindent \tbu{Attack: PnS-TwoParty}
\begin{enumerate}[leftmargin=*]
\item The adversary synchronizes to the first 1 to 0 transition, i.e. start of frame bit (SOF). The transmitting node is the referred to as the primary node.
\item The adversary utilizes the first 19 bits of the header to estimate the expected variation parameters of transition times $(\mu_p, \sigma_p)$. \\

\tbu{Start of the PnS data frame}
\item If the bit value has changed, compute the transition time. Compute the bit triggering the transition by comparing the transition time to a threshold $\tau$, where $\tau$ is a function of $(\mu_p, \sigma_p)$.
\item If bit value has not changed,
\begin{enumerate}
\item If this is corresponding to the first bit (non-inverted), then the possible transitions are $0-0 \rightarrow 0-1$, $0-1 \rightarrow 1-0$, $0-0 \rightarrow 1-0$, and vice versa. If the current voltage level is higher than the previous bit ((b) in Fig. \ref{fig:can_timing_attack}), both nodes transmitted a dominant value for the current bit. Otherwise, if the level decreases, nodes transitioned to a $0-1$ or $1-0$ configuration. Utilize the next bit to compute the current value.
\item If this is corresponding to the second bit, it could have only resulted from a $0-1 \rightarrow 1-0$ transition or vice versa. If a dip is detected at the start of the frame ((a) in Fig. \ref{fig:can_timing_attack}), the primary node was transmitting a dominant bit in the previous frame. Otherwise, if an increase is detected ((c) in \ref{fig:can_timing_attack}), the secondary node was transmitting the $0$.\\

\end{enumerate}

\tbu{Soft synchronization}
\item If the secondary node ever triggers a recessive to dominant transition, resync to the secondary node and switch the roles of the primary and secondary nodes. This is depicted by (r) in Fig. \ref{fig:can_timing_attack}.
\end{enumerate}

\section{Countermeasures for PnS-CAN}
\label{sec:countermeasure}
As discused, the PnS-CAN module is susceptible to a range of characterization attacks due to consistency of the features of the physical signals from individual nodes. While this is good for robustness of the system, the CAN standard allows for significant variation of these properties. Thus, intuitively, we design the countermeasures to add controlled noise to the physical signals, such that signals from different nodes are indistinguishable for an adversary. The goal of the noise addition is to minimize the adversarial advantage between nodes.

Here, we propose mitigations at different levels of abstractions, namely the transceiver hardware level, CAN controller level and the system level. At the hardware and controller level, the goal is to minimize the adversarial advantage between all ECUs either by adding noise or improving cooperation between the nodes. At the system level, the goal is to minimize adversarial advantage by restricting interaction between highly identifiable nodes. It should be noted that the performance of the countermeasures proposed here is highly dependent on the CAN properties. Thus, they are proposed with the goal of minimizing the leakage, rather than provably eliminating it. We leave as an open problem, the design of countermeasures that can formally shown to be effective against the adversaries treated here.

\subsection{Activating Multiple ECUs}
\label{sec:mult_ecu}

The voltage observed by an adversary is a function of strength of the current source and impedance of the network between the observer and the current source. For a non-uniform network, this directly relates to the physical positions of source and the observer. To prevent identification by the adversary, voltage can be varied for each transmission of the dominant bit by using a random number of current sources, located at different points in the network.

\begin{table*}[tb]\footnotesize
\parbox{.45\linewidth}{
\centering
       \caption{$V_{out}$ for dominant tx by multiple nodes}
        \label{tab:mult_dominant_tx}
        \begin{tabular}{ | c c c | c |}
                \hline
                \ndn 1 & \ndn 2 & \ndn 3 & $V_{out}$ \\
                \hline
                0 & 0 & 0 & 2.4230\\
                0 & 0 & 1 & 2.1281\\
                0 & 1 & 0 & 2.1197\\
                0 & 1 & 1 & 1.8208\\
                1 & 0 & 0 & 2.3400\\
                1 & 0 & 1 & 1.7710\\
                1 & 1 & 0 & 1.7629\\
                1 & 1 & 1 & 0.0000\\
                \hline
        \end{tabular}
}\hfill%
\parbox{.45\linewidth}{%
\centering
   \caption{$V_{out}$ for dominant tx by multiple nodes with isolation}
        \label{tab:mult_dominant_tx_cutoff}
        \begin{tabular}{ | c c c | c |}
                \hline
                \ndn 1 & \ndn 2 & \ndn 3 & $V_o$ \\
                \hline
                X & 0 & 0 & 2.5842\\
                X & 0 & 1 & 2.1174\\
                X & 1 & 0 & 2.0923\\
                0 & 0 & X & 2.3159\\
                0 & 1 & X & 1.9647\\
                1 & 0 & X & 2.1493\\
                0 & X & 0 & 2.2957\\
                0 & X & 1 & 1.9599\\
                    1 & X & 0 & 2.1415\\
                \hline
        \end{tabular}
}%
\end{table*}

Our experimental results demonstrate that the observations at a single point due to dominant bits from multiple transceivers is different. Such a phenomenon can also be observed in box (c) of Fig \ref{fig:can_timing_attack}. In Table \ref{tab:mult_dominant_tx}, we enumerate the output at the observer due to simultaneous transmissions from different transceivers connected to the bus. For increased variations, we performed experiments using 3 different transceiver families, i.e. \ndn1 (MCP2551), \ndn 2 (TJA1040) and \ndn 3 (TJA1041). 

Similar to variation in sources, adding a random number of sink nodes may vary the observation by an adversary. In the regular setting, each node on the bus that is not transmitting acts as a sink. Thus, we can vary the number of sinks on the bus by adding the capability of electrical isolation of non-participating nodes. Thus any transceiver with such a capability can be in 3 distinct modes, i.e. transmitting a dominant signal, transmitting a recessive bit (or just passive), or electrically isolated. In Table \ref{tab:mult_dominant_tx_cutoff}, we outline the voltage observations of the adversary for different states of the tri-state nodes. The $`X'$ denotes a transceiver that is electrically isolated. Ideally, with $N$ transceivers, we can obtain $3^N - 2^N$ different voltage levels. The variation of voltage levels with a bit can decrease the probability of its identification. We outline a few architectures to utilize this feature to prevent adversarial information leakage.
\\

\noindent \tb{(Hardware level) Multiple transceivers per controller - } The simplest method to vary the number of active transceivers is by equipping each node with multiple transceivers. For each dominant transmission, the transciver can select any combination of transceiver states to transmit the bit. Similarly for recessive bit transmission, the transceivers can be toggled between the \emph{isolated} and receiving mode. If each controller is equipped with $N$ transceivers with isolation capability, the node can output upto $3^N - 2^N$ voltage levels for the dominant transmission and $2^N - 1$ levels for recessive bit transmission. Thus, for each PnS transmission that is usable as a secret, i.e. one of the nodes transmitting a dominant bit and the other transmitting a recessive bit, the system can cycle between $(3^N - 2^N)(2^N - 1)$ voltage levels. Thus, even a system with $2$ transceivers per controller, the adversary would observe $15$ different voltage levels for each transmission from the node pair.

Though this countermeasure incurs additional cost, one of the advantages of hardware level countermeasures is that they can be implemented independent of the controller. Thus, the frequency of switching between different transceiver states can be higher than the bit rate, effectively causing several noisy variations over the bit period.
\\

\noindent \tb{(Controller level) Passive cooperation - } A lower cost alternative to adding new transceivers is by enabling the nodes connected to the network to cooperate with the transmitting nodes during the PnS-CAN protocol. Each node that does not need to observe signals on the CAN bus can at random time instances, isolate its transceiver (i.e. the `X' state). Note, that since this done only during the PnS-CAN data transmission phase, it does not violate the CAN requirement of nodes listening for critical frames. For nodes with multiple transceivers, this can be implemented without the controller signals, if one of the transceivers always stays connected.
\\

\noindent \tb{(Controller level) Active cooperation - } A node typically monitors the CAN bus by sampling the bus at a high frequency. If it detects a transition from recessive to the dominant state, it can choose to (based on a random choice) actively assist the transmitter by randomly transitioning between the idle and dominant state for the remainder of the bit duration.
\\

\noindent \tb{(System level) Active cooperation - } If the PnS-CAN protocol is executed between a group of nodes, as described in Section \ref{sec:pns_scheme}, all nodes in the chain that precede the currently active nodes are aware of the \emph{expected} transmissions by one of the participants. Thus, active cooperation may be provided by such nodes during the \emph{expected} dominant bit transmission, by synchronously transmitting a $0$. For systems that utilize cooperation at this level, the adversarial advantage between nodes becomes a function of their position within the chain. Thus, the adversarial advantage between a pair of nodes decreases as their position moves further down the chain.

\subsection{(Controller Level) Addition of timing jitter}
As demonstrated by the attack mechanism in Section \ref{sec:timing_attack}, since the nodes synchronize at the beginning of each frame, the transitions for a bit occur within a well defined and distinct time interval, leading to easy classification. Thus, addition of random jitter to each dominant bit transmitted can decrease the probability of identification by the adversary. To achieve this, for each node, \ndn i, we introduce a jitter interval $(t_{N_i}^l, t_{N_i}^h)$, over which the node uniformly adjust its bit start time and correspondingly adjusts the bit duration. The variation of the intervals allows a system designer to minimize the leakage within the jitter tolerance bounds of the system.

\begin{figure}[tb]
\centering
\includegraphics[width=3.5in]{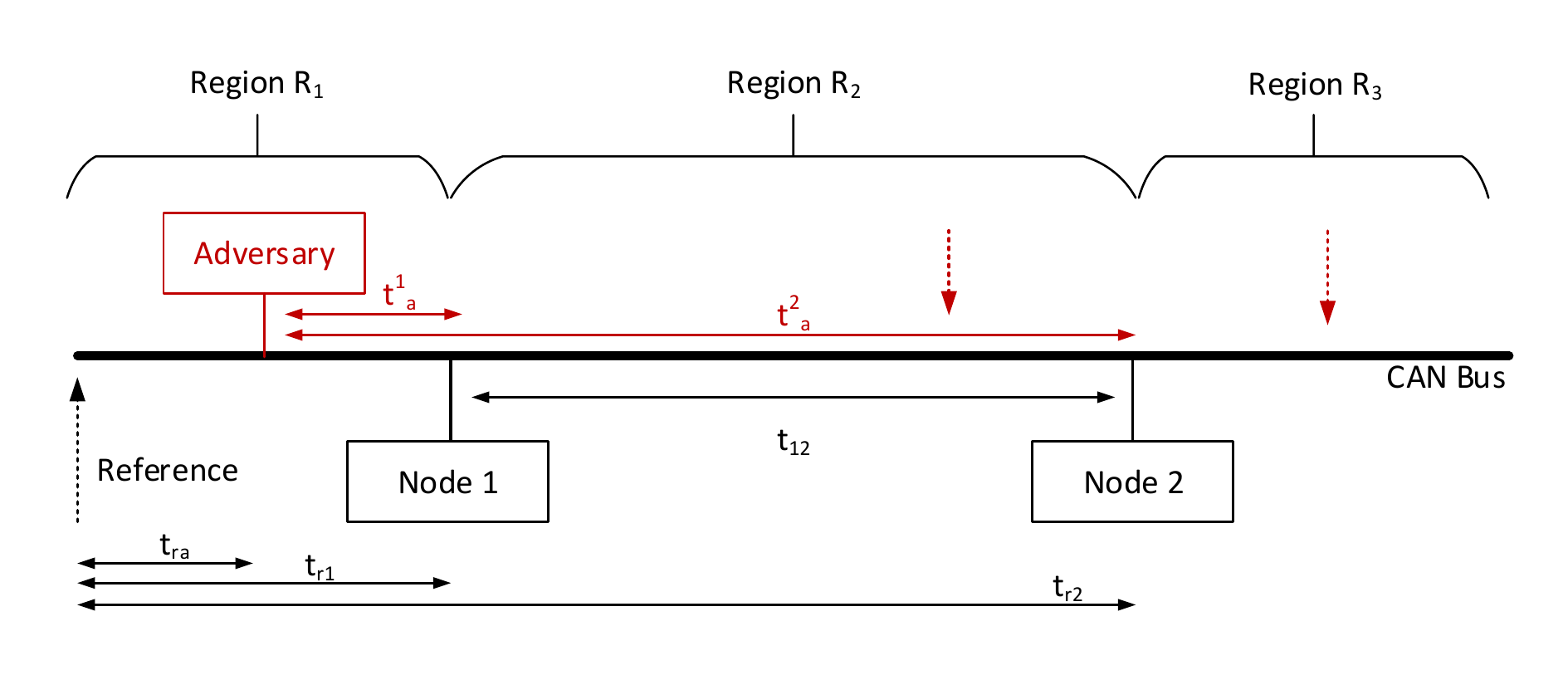}
\caption{Different timing and region parameters of the CAN bus for timing analysis}
\label{fig:network_zones}
\end{figure}

In Fig. \ref{fig:network_zones}, we illustrate the different system parameters that govern the leakage to an adversary for 2 nodes. We define propagation delay from \ndn i to the adversary is $t_a^i, i = 1,2$, and the propagation delay from \ndn 1 to \ndn 2 as $t_{12}$. Assuming that the nodes are synchronized to a common reference point $r$, we represent by $t_{ra}, t_{r1}, t_{r2}$ the offset of the nominal bit start time relative to the common reference. Further, we consider $t_p^2$ to represent the processing time of the bus signal through the circuit of \ndn 2. If the adversary observes transitions from \ndn 1 and \ndn 2, the difference in the observation times of the both the transitions, $\hat{t}_{12}$, can be expressed as%

\begin{align*}
\hat{t}_{12} &= (t_{r2} + t_a^2 + + t_p^2) - (t_{r1} + t_a^1) \\
&= (t_{r2} - t{r1}) +(t_a^2 - t_a^1) + + t_p^2\\
&= (\text{offset between nodes}) + (\text{diff. in prop. delay}) \\
& \quad\quad + \text{(proc time)}.
\end{align*}%

Here, $\hat{t}_{12}$ determines the ability of the adversary to differentiate between the two transmitters. Based on the location of the adversary, the delay varies from $2t_{12} + t_p^2$ in region $R_1$ to $t_p^2$ in region $R_3$. If the nodes utilized the jitter-intervals to determine the start times of the bits, the adversary would observe the bit transition times as samples from the two distributions, over the support sets $I_1 = (t_1^l, t_1^h)$, $I_2 = (t_2^l + \hat{t}_{12}, t_2^h +  \hat{t}_{12})$, for each node \ndn 1, \ndn 2 respectively. The leakage to the adversary is determined by its ability to distinguish between a sample from $I_1$ and $I_2$. Thus the goal of the system designer is to increase the overlap between the two intervals. The nodes minimize the offset as follows,

\begin{enumerate}
\item Prior to PnS-CAN, each node estimates the propagation delay to the communicating node based on a recessive to dominant transition from the other node.
\item The nodes disable soft re-synchronization during the CAN frame transmission. The secondary node \ndn 2 keeps itself synchronized to the primary node \ndn 1 for the duration of the PnS execution.
\item The nodes select a parameter $\alpha$ based on the propagation delay and maximum permissible jitter.
\item The nodes set the jitter intervals as $(t_1^l, t_1^h) = (0, \alpha t_{12})$ and $(t_2^l, t_2^h) = (-\alpha t_{12}, 0)$. The start time of the bit is picked uniformly from the interval.
\end{enumerate}

For practical scenarios, the intervals $I_1$ and $I_2$ cannot be defined to overlap completely. Thus, it is not possible to completely eliminate leakage to the adversary. However, by increasing the overlap, we decrease the adversarial advantage.

\subsection{System Level - Tree-Based Group Key Scheme}
\label{sec:ctr_tree_based}
It is clear that the leakage between different pairs of nodes, varies significantly. An alternative mechanism to reduce adversarial leakage would be to prevent (or minimize) interaction between highly susceptible nodes. Consider the scenario of establishing a key between a group of nodes using the \emph{GroupKey-Linear} protocol in Section \ref{sec:pns_scheme}. Intuitively, the order of the node communication can be modified to prevent leakage prone nodes from communicating. First we consider the following modification to the \emph{GroupKey-Linear}, for organization of the nodes in a tree-like structure.
\\

\noindent \tbu{GroupKey-Tree:}
\begin{enumerate}
\item The group communication order $G_K = \{N_1, N_2, \ldots, N_k\}$ is broadcast to all participants by the gateway ($GW$).
\item First, $N_1$ performs PnS with $N_2$ to establish a key. Then $N_2$ utilizes the result to perform PnS with any $N_i,~i \leq 2$ and updates the key. Proceeding forward, $N_j$ will performs PnS with any $n_i,~i < j$.
\item The process continues till $n_k$ is reached and the group key is generated.
\end{enumerate}

Functionally, the \emph{GroupKey-Tree} protocol is equivalent to \emph{GroupKey-Linear}. The \emph{GroupKey-Tree} scheme presented here should not be confused with the tree-based scheme presented by the original authors in \cite{jain_jorge_2016_ches}, as that utilizes a binary balanced tree. We have no such constraints here. It should be noted that for both \emph{GroupKey-Tree} and \emph{GroupKey-Linear}, the number of interactions between the nodes are the same. Thus \emph{GroupKey-Tree} introduces no additional communication overhead.
 
Successive interactions leak information to an adversary, and such leakage is dependent upon the participants. Thus intuitively, the each node in $G_K$, instead of communicating with a fixed single node, can select the node from a pool that minimizes adversarial leakage. We utilize the adversarial advantage described in Section \ref{sec:gen_model} to establish a leakage-aware node order for group key generation as follows,
\\

\noindent \tbu{1. Mapping} - First, the $GW$ maps the network and the adversarial advantage between all pairs of nodes $\{1, 2, \ldots, M\}$. The adversarial advantage can be represented as weights of the edges of a fully connected graph $K_M$, where the ECUs represent the vertices. Thus, the $GW$ learns the weighted graph. This can be performed by several methods, i.e determined offline during system installation and programmed into the $GW$, or using observing network data to  dynamically determine classification functions to compute the adversarial advantage.
\\

\noindent \tbu{2. Ordering} - For any given group $G_K$, the $GW$, using the graph obtained during mapping, reorders the nodes such that the overall leakage is minimized. For the \emph{GroupKey-Tree} scheme, the node order takes the form of a tree. The criteria of minimization can be varied based on the countermeasures for pairwise PnS and adversarial restrictions.

We define by $K_M[G_K]$, the subgraph induced in $K_M$ by $G_K$. Thus, for key generation using \emph{GroupKey-Tree}, the process of defining the node order using $K_M[G_K]$ is tantamount to finding the minimum spanning tree over the graph. Define $\mc T$ as the set of all possible spanning trees of the graph $K_M[G_K]$. We define the node ordering problem as finding $T^* \in \mc T$ that solves,
\begin{equation}
\min_{T \in \mc T} \max_{(i,j) \in T} d_{i,j},
\end{equation}
where $(i,j)$ is the edge in the tree and $d_{i,j}$ is the adversarial advantage for nodes $i$ and $j$. This minimizes the maximum cost path in the tree. For a group key scheme over the tree, leakage at any step would leak the entire key. Intuitively, we want to ensure that the weakest path in the scheme does not leak sufficient information to the adversary. Computationally, this requires the adversary to compute the min-max spanning tree over $K_M[G_K]$. Since every minimum spanning tree is also a min-max spanning tree, the solutions that minimize total leakage also minimize the max-link leakage. Thus it is both computationally and performance optimal to simply compute the minimum spanning tree over the subgraph. Several well known algorithms are present in literature for such optimization, e.g. Prim's algorithm \cite{prim_57_mst}, Kruskal's algorithm \cite{kruskal_56_mst}.
\\

\noindent \tbu{3. Rank assignment} - Once the order of communication has been determined by the $GW$, can be communicated to the members of the group. As discussed in Section \ref{sec:gen_model}, the lack of knowledge about the node order would force the adversary to use a generic classification function $\mc D_{0, \bar{0}}$, thus effectively decreasing the adversarial advantage. Thus, the $GW$ can use `privacy preserving' methods to distribute the node order. This can be achieved by several mechanisms based on the adversarial assumptions and pre-shared knowledge between the $GW$ and the nodes.

As an example, consider a system using the authenticated variant of the PnS-CAN scheme, \cite{jain_jorge_2016_ches}. Here, the $GW$ shares a secret (cryptographic key) with each node. Thus a node can utilize this key to mask the identity of each node, and hence the overall order of the nodes. Consider $f(\cdot, \cdot)$ to be a cryptographic pseudorandom function (PRF). Thus, the order of the group $\{N_{O_1}, \ldots, N_{O_K}\}$ may be communicated as
\[ r || f(k_{N_{O_1}}, r) || f(k_{N_{O_2}}, r) || \ldots || f(k_{N_{O_K}}, r), \] where $r \leftarrow \{0,1\}^n$ represents a nonce.

\section{Discussion}
\label{sec:discussion}
The goal of the discussion in this paper is twofold. Firstly, it demonstrates that even simple physical features can be utilized to compromise the PnS-CAN scheme. Secondly it illustrates simple techniques that can be utilized to modify the physical characteristics and reduce the information leakage. We emphasize that neither of these investigations is intended to represent the comprehensive attack vectors or defense mechanisms.

For example, given the robustness of the CAN protocol, one can imagine that there are a series of fundamental component variations that can be utilized to add noise to the characteristics discussed here. Similarly, future results in classification and learning theory can be utilized to improve the performance of the adversary beyond what has been discussed here.

There are several open questions that remain with the countermeasures discussed here. For example, there are fundamental trade-offs that can be investigated between the countermeasures. As an example, addition sufficient jitter to the bit timing for effectiveness, while maintaining robust sampling between the farthest nodes may require operation of the system at a lower speed. However, this improves the observability of the adversary and significantly impacts transients (and any associated countermeasures). Further, improved countermeasures can be developed by identifying correlations between the adversarial advantage and physical features. While we have studied certain empirical dependencies, a comprehensive analysis can yield better insight into identification of dominant physical characteristics that can be modified.

For our attacks and countermeasures, we utilized a simple laboratory setup to emulate the CAN system. A real deployment of the system is subject to far more noisy conditions and many subtle electrical effects, e.g. transmission line phenomenon, that are not accurately in our setup. Thus even though this work serves as a proof-of-concept, the investigation of these attacks and countermeasures can yield different results in a real system, and warrants further investigation.

The countermeasures discussed here span several implantation layers and components. While simple hardware and controller changes can uniformly improve the system security for all the nodes, they require changes to the CAN transceivers or controllers, which may not be very desirable in a real system.
Further, for deployment of such systems in highly regulated environments, such as automotive, the new changes would have to be thoroughly tested to not impact the system reliability. This can significantly increase the cost of implementing such countermeasures.

On the other hand, system level changes can be easily implemented, but have impact on leakage for limited scenarios. In a practical system, the designer would deploy several additional mechanisms in addition to the countermeasures proposed here.

For example, there may be system level mechanisms that limit the data gathering ability of an adversary. Since the performance of the distinguishing functions used by an adversary depends on the training data available to it, lack of data can reduce the leakage. Similarly, the system designer may electrically isolate common access mechanisms, such as the OBD-II port to limit the observation of physical characteristics. As discussed briefly, the lack of apriori knowledge about the participants of the protocol can reduce adversary performance by limiting the distinguishing functions that can be used. Such masking techniques could be implemented by designer at a system level in addition to the proposed countermeasures.

\subsection{Conclusion}
\label{sec:conclusion}
While the PnS-CAN scheme is a promising mechanism for generating group keys, it is highly susceptible to probing attacks by even simple adversaries. We presented a general methodology for attacking the system and demonstrated a simple timing based attack. We discussed several sources of physical identifiers that can be used by an adversary. We proposed several mechanisms to mask these identifiers at the hardware, CAN controller and system level. This work serves as a proof-of-concept for the existence of attacks on the PnS-CAN schemes and the the ability of the system designer to prevent them.

\begin{appendices}

\section{Node Distance}
\label{sec:node_dist}
To characterize the difference in leakage between different nodes, we define the adversarial advantage for interaction between two nodes $(N_i, N_j)$, as probability of successfully distinguishing between them, i.e. identify the transmitting node, based on the bus observations. For practical scenarios, we can define the adversarial advantage based on the performance of the trained classification functions defined in Section \ref{sec:gen_model}, applied to empirical data. We define by $x^{N_j}_i$, the samples of transmissions of bit $i$ from node $N_j$. Thus we define the adversarial advantage for nodes $(N_i,N_j)$ as, 
\begin{align}
d_{N_i,N_j} & = \frac{1}{4} \Big(  Pr\left(\mc D_{i, j}(x^{N_1}_1, x^{N_1}_2) = (0, 0)\right) \nonumber \\
& +  Pr\left(\mc D_{i, j}(x^{N_1}_1, x^{N_2}_2) = (0, \bar{0})\right) \nonumber \\
& +  Pr\left(\mc D_{i, j}(x^{N_2}_1, x^{N_1}_2) = (0, \bar{0})\right) \nonumber \\
& +  Pr\left(\mc D_{i, j}(x^{N_2}_1, x^{N_2}_2) = (0, 0)\right)
\Big),
\label{eqn:dist_prob}
\end{align}
\noindent where $Pr$ denotes the probability over all random choices made in the decision function. 

In \eqref{eqn:dist_prob}, we assume each sequence is equi-probable, which is typically the case for the PnS-CAN protocol. For scenarios where information about the participating node is not available, we define a generalized adversarial advantage $dg_{i,j}$ by using the generalized decision function $D_{0, \bar{0}}$ in place of $D_{i, j}$. Clearly, $dg_{i,j} \leq d_{i,j}$ for all nodes. 

Such a notion of distance can be used to quantitatively measure the impact of the countermeasures implemented by the system designer. Alternatively, this can provide a numerical measure of the adversarial strength for any optimization that the designer may choose to perform.
\end{appendices}

\bibliographystyle{IEEEtran}

\end{document}